\documentstyle[eqsecnum,aps]{revtex}

\def\?{\phantom{1}}

\begin{document}
\title{Metal-Nonmetal Changeover in Pyrochlore Iridates}

\author{D. Yanagishima}
\address{
Department of Physics, Kyoto University, Kyoto 606-8502, Japan}

\author{Y. Maeno}
\address{
Department of Physics and Kyoto University International Innovation Center, 
Kyoto University, Kyoto 606-8502, Japan\\
CREST, Japan Science and Technology Corporation}

\date{\today}
\maketitle
\begin{abstract}
We report the low temperature properties of a new series of 
pyrochlore iridates \it{R}\rm $_2$Ir$_2$O$_7$ (\it{R} \rm{=} rare-earth 
elements). We found that the compounds with \it{R} \rm{=} Pr, Sm, Nd, 
and Eu exhibit metallic conductivity, whereas those with smaller 
rare-earth ions are nonmetallic. Such metal-nonmetal changeover is 
attributable to the importance of electron correlation among the Ir 5d 
electrons. Reflecting the geometrical frustration 
in the pyrochlore lattice, the rare-earth moments do not exhibit magnetic 
ordering to temperatures well below the antiferromagnetic Weiss temperature. 
We have not found any sign of superconductivity down to 0.3 K 
in these compounds. 
\end{abstract}
\twocolumn
\quad
The ground state degeneracy in geometrically frustrated systems has been a fascinating reserch subject \cite{spin}.
Recently, the pyrochlore compounds represented by the composition $A_{2}B_{2}$O$_7$ have been extensively investigated as the model systems.
The pyrochlore structure consists of a three-dimensional network of corner-sharing tetrahedra both $A_{2}$O and $B_2$O$_6$ sublattices.
One of the remarkable phenomena associated with the frustrated magnetism of the $A$ site is the "spin ice" behavior \cite{HarrisNature,Harris97} with residual zero-point entropy observed in the insulating pyrochlore Dy$_2$Ti$_2$O$_7$ \cite{Ramirez}. 
It is the nearest-neighbor dipolar interaction, dominating over the antiferromagnetic superexchange interaction, that provides ferromagnetic interaction necessary for the frustration \cite{Hertog}.
Moreover, it is  crucial to include the long-range nature of dipolar interaction to describe the details of the properties \cite{dipolarSI};
thus, the description is termed a "dipolar spin ice" model.
This raises an important question whether the spin ice and related behavior is realized in the presence of other kinds of magnetic interaction with finite range, such as RKKY interaction in metallic pyrochlores.

\quad
The physics associated with the d electrons of the $B$-site ions is without doubt equally interesting. 
Adjusting the d-electron bandwidth by varying the $A$-site ions with different sizes leads to metal-insulator changeover in the molybdenum series \it R\rm $_2$Mo$_2$O$_7$ \cite{Ali}.
In these systems, ferromagnetic ordering of itinerant moment, originating from Mo 4d electrons with 4d$^2$ (spin "$S$ = 1") configuration, occurs at relatively high temperatures of 50 to 80 K.
For the metallic ferromagnets, an intimate magnetic interplay between the magnetically ordered Mo and rare-earth moments results in an unusual Hall effect specific to the geometrically frustrated system \cite{Taguchi,Katsufuji,Sato}.
At this stage of the research on the geometrically frustrated magnetic systems, it is very important to explore other pyrochlore systems composed of elements with qualitatively different physical properties. 
For example, it is theoretically anticipated that a spin liquid state be realized for Heisenberg $S$ = 1/2 spins on the pyrochlore lattice \cite{Canals}.
Very recently, superconductivity of the pyrochlore Cd$_2$Re$_2$O$_7$ 
with \it{T}$_c$ \rm= 1.1 K has been discovered independently by Hiroi 
\it{et al.} \rm and by Yoshimura \it{et al.}\rm \cite{Cd2Re2O7}. 
This discovery further urges the study of 
low-temperature properties of a variety of pyrochlore compounds. 

\quad
We report here the resistivity, magnetic susceptibility, as well as the specific heat of a new series of pyrochlore componds based on 
iridium, \it{R}\rm $_2$Ir$_2$O$_7$ (\it R \rm = rare-earth elements). 
The tetravalent Ir containing five 5d electrons in the low-spin $S$ = 1/2 configuration indeed brings about qualitatively different properties from those of componds based on Mo and Ru with \it S \rm = 1 configuration \cite{Ru227}.
Pyrochlore iridates with \it R \rm = Pr, Nd, Sm, and Eu show metallic conductivity without any indication of magnetic ordering associated with Ir spins.
The metal-nonmetal changeover occurring with smaller rare-earth ions is attributable to the electron correlation among the Ir 5d electrons.
This new system may prove very useful in testing the predictions of the \it S \rm = 1/2 Heisenberg spins on a pyrochlore lattice \cite{Tsun,Fuji}.

\quad
We have succeeded in synthesizing the new series \it R\rm $_2$Ir$_2$O$_7$ with various lanthanide elements as well as Y for \it R\rm . 
Syntheses for \it R \rm = Eu \cite{Bouchard} and Gd \cite{Gd} were previously reported, but the low temperature properties had not been known.
We synthesized high-quality polycrystalline samples of \it R\rm $_2$Ir$_2$O$_7$ by a standard solid-state synthesis technique.
The starting materials|rare earth oxide (Pr$_6$O$_{11}$, Tb$_4$O$_7$, and \it{R}'$_{\rm{2}}$\rm{O}$_3$ for \it{R}' \rm = Nd, Sm, Eu, Gd, Dy, Ho, Er, Yb, Y; 99.99\%) powder and IrO$_2$ (99.9\%) powder|were mixed either in the stoichiometric ratio or in the moler ratio of 1:1.1 and heated in air at temperatures ranging from 790$^{\circ}$C and 950$^{\circ}$C for one week with several intermediate gridings.
Depending on the morphology of the IrO$_2$ powder used, higher temperatures up to 1100$^{\circ}$C were also needed.
The phase purity of the products were confirmed by x-ray diffraction.
We have tried but not suceeded in growing single crystals by a floating-zone technique.

\quad
We measured the resistivity by a standard four-probe method between 4.2 K (50 mK for \it{R} \rm = Pr) and 300 K, the ac susceptibility by a mutual-inductance method between 0.3 K and 4.2 K, the dc magnetization with a commercial SQUID magnetometer(Quantum Design, MPMS) between 1.8 K and 300 K, and the specific heat by a relaxation method(Quantum Design, PPMS) between 1.8 K (0.4 K for \it{R} \rm = Pr, Eu, and Dy) and 300 K. 
We have not clarified the physical properties of Er$_2$Ir$_2$O$_7$, since this compound did not sinter well.

\quad
We illustrate the resistivity in Fig. 1.
Although only qualitative significance is drawn from these policrystalline data, the temperature dependence as well as the magnitude clearly indicate that the compounds with \it{R} \rm = Pr, Nd, Sm, and Eu are metallic,
whereas those with Gd, Tb, Dy, Ho, and Yb are nonmetallic.
The metal-nonmetal boundary is between Eu and Gd in the periodic table, and is well correlated with the ionic radius \cite{Shannon} and the cubic lattice parameter, as shown in the inset of Fig. 1.
The lattice parameter increases linearly with the ionic radius of \it R\rm $^{3+}$ for the nonmetallic compounds. 
For the metallic compounds, however, it is somewhat smaller than the extrapolation from the nonmetallic side.
A similar tendency is reported in the Mo-based pyrochlores, for which the boundary between ferromagnetic metal and spin-glass insulator is between Gd and Tb \cite{Ali}.
Such metal-nonmetal transition is attributed to the effect of correlation among the 4d electrons \cite{Taguchi}.

\quad
We searched for possible superconductivity by ac susceptibility measurements down to 0.3 K with a frequency of 1kHz and an ac field of 0.1 mT, but did not find any evidence as shown in Fig. 2. The peak at 1.35 K for \it R \rm = Nd is probably due to antiferromgnetic ordering of the Nd moments as discussed below. 

\quad
We show in Fig. 3 the inverse dc magnetization under the field of 1 T.
The magnetic properties are characterized by the local moment of the trivalent rare-earth ions and the itinerant/localized tetravalent Ir with 5d$^{5}$ configuration in the triply degenerated t$_{2 \rm g}$ orbits.
Except for Sm$_2$Ir$_2$O$_7$ and Eu$_2$Ir$_2$O$_7$, the susceptibility $\chi = M/H$ is well fitted with the Curie-Weiss law: $\chi = \chi _{\rm 0} + C/(T - \theta _{\rm CW})$.
The parameters obtained by fitting between 100 and 300 K are summarized in 
Table I.
The effective moment \it p\rm $_{\rm eff}$, derived from the Curie constant \it C\rm , agrees well with the expectation for \it R\rm $^{3+}$ ions, indicating the localized nature of the trivalent rare-earth ions.
The Weiss temperatures $\theta _{\rm CW}$ are all antiferromagnetic.
It should be noted that although Dy$^{3+}$ and Ho$^{3+}$ are known to yield \it{ferromagnetic }\rm $\theta _{\rm CW}$ in \it R\rm $_2$Ti$_2$O$_7$, both exhibit antiferromagnetic $\theta _{\rm CW}$ in \it R\rm $_2$Ir$_2$O$_7$. This diffrence suggests the involvement of the Ir spins in determining the interaction between the rare-earth moments.
The small and only-weakly temperature-dependent susceptibilities of the Sm and Eu compounds are quantitatively consistent with the main contribution from the Van Vleck susceptibility. 
For these compounds, there is no sign of the additional \it S \rm = 1/2 
local moments of the Ir ions, consistent with the Pauli suceptibility expected from the observed electronic specific heat (below).
For the nonmetallic compounds, the contribuion of the Ir local moment could not be precisely extracted because of the large contribution from the rare-earth moments. 
For Y$_2$Ir$_2$O$_7$ with nonmagnetic Y$^{3+}$, ferromagnetic contribution with the Curie temperature \it T\rm$_{\rm C}$ = 170 K, possibly due to impurities or grain bowndaries, hampered the determination of the Ir contribution.
\quad
Figure 4 shows the specific heat devided by temperature, $C/T$.
Indication of long-range magnetic ordering is found only in \it R \rm = Yb at 2.2 K; the others exhibit no indication of phase transition at least down to 1.8 K, although the the gradual increase of $C/T$ is attributable to the freezing of the moments. 
As in Fig. 2, the ac susceptibility measurements extended down to 0.3 K indicates additional magnetic ordering in \it R \rm = Nd. Thus, \it R\rm $_2$Ir$_2$O$_7$ remains paramagnetic at temperatures much lower than the respective Weiss temperatures, most probably reflecting the geometrical frustration on the pyrochlore lattice.

\quad
The specific heat of Pr$_2$Ir$_2$O$_7$ is qualitatively different from those of the other compounds: $C/T$ exhibits a broad peak with the maximum at 4.6 K.
This compound remains as metallic at least down to 50 mK and paramagnetic down to at least to 0.3 K.
Since the effective moment \it p\rm $\rm _{eff}$ obtained from the Curie-Weiss fitting at high temperatures is consistent with that of a localized Pr$^{3+}$ ion, the itenerant character is attributable to electrons originating from Ir$^4+$.
Owing to the crystal-field splitting of 4f$^2$ configuration of Pr$^{3+}$ ion in the expected crystal symmetry of $D _{3d}$, we expect the low-lying states to be composed of a singlet and a magnetic doublet.
In order to examine the entropy associated with the specific heat peak, we subtracted the estimated contribution from the lattice and the itinerant electrons by using the data of metallic Eu$_2$Ir$_2$O$_7$, for which Eu$^{3+}$ ion has no effective ground-state moment.
The specific heat of Eu$_2$Ir$_2$O$_7$ itself yields the Debye temperature of ${\mathit\Theta}$$_{\rm D}$ = 420 K, and the electronic coeficient of ${\gamma}$ = 7 \rm{mJ/K}$^2$ mol-Ir.
The resulting magnetic entropy is 4.8 J/K mol-Pr at 25 K, not reaching the expected value of $R$ ln3 = 9.14 J/K mol; it is even smaller than $R$ ln2 = 5.76 J/K mol.
A simple model based on either a singlet or a doublet ground state cannot account for the observed behavior.
Such small entropy, however, is consistent with the level scheme with a frustrated ground-state doublet and an excited singlet. In fact, $C/T$ starts to increase below 0.5 K as shown in Fig. 4, suggesting the release of the entropy associated with the doublet at lower temperatures.
Therefore, future studies extended to lower temperatures is needed to identify the crystal-field scheme.

\quad
We noted above that the Weiss temperature is antiferromagnetic even for \it{R }\rm = Dy and Ho. 
The dominant interactions for insulating pyrocholore magnets are the superexchange and the dipolar interactions. The dipolar interaction is ferromagnetic and evaluated as ${D_{\rm nn} = 5\mu_{\rm 0}{p _{\rm eff}}^2/12\pi{r_{\rm nn}}^3}$ with \it r$_{\rm nn}$ \rm being the distance between the \it R$^{\rm 3+}$ \rm ions.
Using \it r$_{\rm nn}$ \rm = 3.60(4) \AA ~deduced from the lattice parameter given in Fig. 2, we estimate it to be about 2.51 K, not very different from that of Dy$_2$Ti$_2$O$_7$, for which $D_{\rm nn}$\ =\ 2.57 K combined with the antiferromagnetic nearest neighbor exchange interaction of $J_{\rm nn}$\ = \ $-1.24$ K is used to account for the positive Weiss temperature \cite{Hertog}. 
The observed $\theta _{\rm{CW}} = - 3.5$ K suggests a leading contribution of the exchange interaction involving the Ir spins.
Moreover, the negative $\theta _{\rm{CW}}$ suggests that the ice-rule, necessary for the strong geometrical frustration, is no longer governing in Dy$_2$Ir$_2$O$_7$.
In order to examine the effect of sign change of the effective magnetic interaction on the low-temperature entropy, we compare the entropy of Dy$_2$Ir$_2$O$_7$ and Dy$_2$Ti$_2$O$_7$ in Fig. 5. 
The entropy is obtained by integrating $C/T$ over temperature, after
the phonon contribution estimated from the respective data of the Eu-based compounds have been subtracted.
Clearly, the zero-point entropy of 1/2 ln(3/2) present in the spin-ice compound Dy$_2$Ti$_2$O$_7$ no longer exist in Dy$_2$Ir$_2$O$_7$, and the full spin entropy of $R$ ln2 is recovered above 20 K.

\quad
The inset of Fig. 5 shows $C/T$ of Dy$_2$Ir$_2$O$_7$ below 5 K.
The temperature dependence of the main part is typical of the spin freezing without long-range ordering. 
However, there is an additional sharp peak at 1.2 K with little thermal hysterisis. Studies with microscopic probes are under way to help clarifing the origin of this peak.

\quad
In summary, we have succeeded in synthesizing the new series of pyrochlore compounds \it{R}\rm $_2$Ir$_2$O$_7$.
The metal-nonmetal changeover indicates the importance of electron correlations among the Ir 5d electrons.
In the metallic compounds, we found that Ir 5d electrons exhibit a respectable mass enhancement, but we have not found any indication of superconductivity down to 0.3 K.
In the nonmetallic members, we could not determine the magnetic state of the Ir \it S \rm = 1/2 spins in the present study,
owing to a large magnetic contribution from the rare-eath moments.
We can at least state that strong ferromagnetic ordering, like the one observed in Lu$_2$V$_2$O$_7$ originating from $S = 1/2$ of V$^{4+}$ ions, \cite{Shin-ike} is absent in \it{R}\rm $_2$Ir$_2$O$_7$. 
The difference in the number of d electrons, one for V$^{4+}$ and 5 for Ir$^{4+}$, must be making an essential difference in their magnetic ground states.
The spin-ice behavior is absent in the  nonmetallic Dy compound, since the near-neighbor interaction is antiferromagnetic.
Such sign change of the interaction suggests the important involvement of the Ir spins, absent in the Ti pyrochlores.

\quad
The authors acknowledge H. Fukazawa and R. Higashinaka for their contribution, N. Kikugawa, T. Ishiguro and H. Yaguchi for their support, and M.J.P. Gingras, S. Fujimoto, H. Tsunetsugu, and K. Ueda for usuful discusions. 

\vspace{5mm}
\begin{table}
\begin{center}
\caption{Curie-Weiss fitting parameters for \it R\rm $_2$Ir$_2$O$_7$.}
\label{t1}
\begin{tabular}{cccccccc} 
\it R & Pr & Nd & Gd & Tb & Dy & Ho & Yb \\ \hline
$p_{\rm eff}$ (obs.)  & 3.00 & 3.20 & 8.18 & 9.62 & 10.1\? & 10.3\? & 3.55 \\ 
$p_{\rm eff}$ (calc.) & 3.58 & 3.62 & 7.94 & 9.72 & 10.63 & 10.58 & 4.54 \\ \hline
$\theta_{\rm CW}$ (K) & $-10$ & $-19$ & $-7.8$ & $-14$ & $-3.5$ & $-0.83$ & $-9.3$ \\  
\end{tabular}
\end{center}
\end{table}
\begin{figure}
\caption{Resistivity of polycrystalline \it R\rm $_2$Ir$_2$O$_7$. The compounds with \it 
R \rm = Pr, Nd, Sm, and Eu are metallic and those with Gd, Tb, Dy, Ho, 
and Yb are nonmetallic. Inset: The relations between the ionic radius of 
\it R\rm $^{3+}$ and the observed cubic lattice parameter. The metallicity is 
intimately correlated with the cell size.}
\end{figure}
\begin{figure}
\caption{Low-temperature ac susceptibility of metallic \it R\rm $_2$Ir$_2$O$_7$
with \it R \rm = Pr, Nd, Sm, and Eu.
The superconducting transition of In metal is also shown to indicate the 
approximate magnitude of the diamagnetic signal expected for superconductivity.
The vertical scale of each data is offset for clarity.
Evidence for a magnetic ordering at 1.35 K, most probably the antiferromagnetic
odering of Nd moments, is seen for the Nd compound.}
\end{figure}
\begin{figure}
\caption{Inverse dc magnetization at low temperatures.
The susceptibility is well fitted with the Curie-Weiss law except for 
$R$ = Sm and Eu, for which contribution from the Van Vleck term is dominant.}
\end{figure}
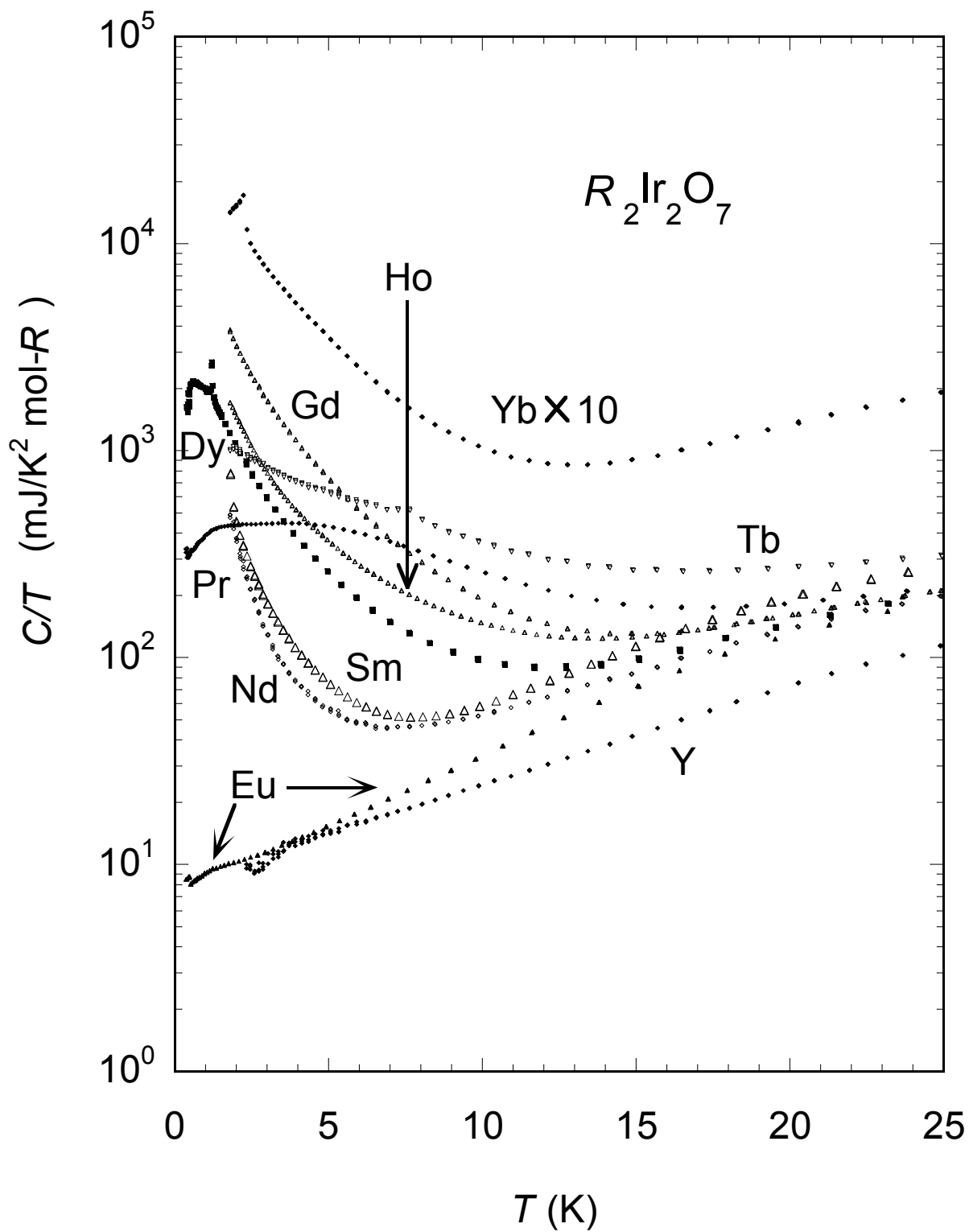
\begin{figure}
\caption{Specific heat over temperature of $R$$_2$Ir$_2$O$_7$. 
The Pr compound exhibits qualitatively different low-temperature behavior 
among the ones with rare-earth moments.
The data for \it R \rm = Yb, exhibiting the anomaly associated with a magnetic ordering at 2.2 K, is offset for clarity.}
\end{figure}
\begin{figure}
\caption{Comparison of the entropies of 
Dy$_2$Ir$_2$O$_7$ and the spin-ice Dy$_2$Ti$_2$O$_7$. The phonon contributions,
estimated from the data of the respective Eu compounds, have been substracted.
Dy$_2$Ir$_2$O$_7$ does not exhibit the zero-point entropy. 
Inset: the specific heat over temperature of Dy$_2$Ir$_2$O$_7$ below 5 K.
The phonon subtraction has not been made, but it has a negligible effect 
in this temperature range.}
\end{figure}

\end{document}